\documentclass{mem}
\usepackage{txfonts}\usepackage{balance}
\usepackage{graphicx}
\usepackage[a4paper]{hyperref}
\idline{75}{282}
\begin{document}
\def\teff{$T\rm_{eff }$}
\def\kms{$\mathrm {km s}^{-1}$}

\title{
Multiple stellar populations in the Globular Clusters NGC1851 and
NGC6656 (M22).
}

   \subtitle{}

\author{
A. \,P. \, Milone\inst{1}, 
G. \, Piotto\inst{1}, 
L. \,R. \, Bedin\inst{2}, 
A. \,F. \, Marino\inst{3}, 
Y. \, Momany\inst{4}, 
\and S. \, Villanova\inst{5}
          }

  \offprints{A. P. Milone}

\institute{
Dipartimento  di   Astronomia,  Universit\`a  di
  Padova, Vicolo dell'Osservatorio 3, Padova, I-35122, Italy
\and
Space Telescope Science Institute, 3700 San Martin Drive,
Baltimore, MD 21218, USA
\and
Max Plank Institute for Astrophysics, Postfach 1317, 85741, Garching,
Germany 
\and
European Southern Observatory, Alonso de Cordova 3107, Vitacura, Santiago, Chile 
\and 
Grupo de Astronomia, Departamento de Física, Casilla 160, 
              Universidad de Concepci\'on, Chile
}

\authorrunning{Milone et al. }

\titlerunning{Multiple stellar populations in NGC1851 and M22}

\abstract{
In the last years, photometric and spectroscopic evidence has demonstrated that
many, maybe all the Globular Clusters (GC) host multiple stellar
populations. 
%
%
High-resolution spectroscopy has established that, while most GCs are
mono-metallic with no significant abundance spread in $s$-elements, in
all the globulars studied to date the presence of different stellar
generation is inferred by the Na-O and the C-N anticorrelations.  

In this context, NGC 1851 and NGC 6656 are among the most intriguing
clusters. Contrary to the majority of GCs, they host two groups of
stars with different s-elements abundance that are clearly associated
to the two distinct sub-giant and red-giant branches detected in their
color-magnitude diagrams (CMD).  
In the case of NGC 6656 s-rich stars are also enriched in iron and calcium.
Each $s$-element group exhibits its own Na-O and C-N anticorrelations thus
 indicating the presence of sub-populations and suggesting
 that the parent clusters have experienced a very complex
 star-formation history.  
In this paper we summarize the properties of multiple populations in
NGC 1851 and NGC 6656. 
}
\maketitle{}

\section{Introduction}
In recent years,
an increasing number of photometric and spectroscopic observational
evidence have shattered the paradigm of globulars as the prototype
of single, simple stellar populations (see Piotto \ 2009 for a recent
review).   

Spectroscopic studies have demonstrated that most GCs are
mono-metallic with no detectable spread in their iron content
 and also $s$-process elements do not exhibit large star-to-star
 variations in the majority of globulars  (e.g. Carretta et al.\ 2009a, 
D'Orazi et  al.\ 2010 and references therein). On the contrary, in
 all the clusters 
 studied to date with large stellar samples, have been detected
 star-to-star variations in the light elements C, N, O, Na, and Al
 (e.g. Carretta et al.\ 2009b, Pancino et  al.\ 2010 and references therein).
These variations are related to correlations and anticorrelations,
which indicate the occurrence of high temperature hydrogen-burning
processes which include CNO, NeNa, MgAl cycles and cannot occur in
presently observed low mass GC stars.    

Today it is widely accepted that the observed light-elements variations
provide strong support to the presence of multiple stellar population
in GCs with the second generations formed from the material polluted
by a first generation of stars. On the contrary the debate on the
nature of possible polluters is still open (e.g. D'Antona \& Caloi
al.\ 2004, Decressin et al.\ 2007).

While  abundance variations are well known since the early
sixties, it was only the recent spectacular discovery of multiple
sequences in the CMD of several GCs that provide an un-controversial
prove of the presence of multiple stellar populations in GCs and
brought new interest and excitement in GCs research (e.g. Piotto et
al.\ 2007).
Photometric clues, often easy to detect simply by inspection of
high-accuracy CMDs, arise in form of multiple main sequences (MS,
Bedin et al.\ 2004, Piotto et al.\ 2007, Milone et al.\ 2010), split
sub-giant branch (SGB, Milone et al.\ 2008, Anderson et al.\ 2009,
Piotto \ 2009), and multiple red-giant branch (RGB, Marino et
al.\ 2008, Yong et al.\ 2008, Lee et al.\ 2009).

Apparently the main properties of stellar generations, like the relative
fraction of stars, their location in the CMD, the spatial
distribution, and the chemical behavior differ from 
cluster to cluster. 
However, it is clear that some groups of clusters share all similar
properties: 
\begin{enumerate}
\item 
Na-O anticorrelation and C-N correlations are presents in all the
GCs studied to date.  When stars with available Na and O abundances 
have been identified in the {\it U} vs. ({\it U-B}) CMD, it was found that 
the group of Na-poor (O-rich) stars are spread on the blue side
 of the RGB, while the Na-rich (O-poor) population define a narrow 
 sequence on the red RGB (e.g. Marino et al.\ 2008).
\item
In few cases the MS morphology supports the presence of stellar populations with
different helium. 
NGC 2808 has three distinct MSs (Piotto et al.\ 2007) possibly
associated to three stellar 
population with primordial helium and with Y$\sim$0.33 and Y$\sim$0.40 (D'Antona et al.\ 2005). A
spread MS have been detected  
in NGC 104 (47 Tuc, Anderson et al.\ 2009) and  NGC 6752 (Milone et
al.\ 2010) where there is also some hint of a MS split.  
\item
In some GCs, like NGC 104, NGC 1851, NGC 5286, NGC 6388,
  and NGC 6656 there is a double SGB (Milone et al.\ 2008,  Piotto
\ 2008,  Anderson et al.\ 2009) associated to two stellar
groups with either a difference in age by $\sim$1 Gyr or with
almost the same age but a significant difference in their overall
C+N+O content (Cassisi et al.\ 2008, Ventura et al.\ 2009, Di Criscienzo et al.\ 2010).  
\item
Finally there is the `extreme' case of
 $\omega$ Centauri with either multiple MSs (e.g. Anderson\ 1997, Bedin et al.\ 2004), 
multiple SGBs (e.g. Sollima et al.\ 2005, Villanova et al.\ 2007), multiple RGBs (Lee et al.\ 1999, Pancino et al.\ 2000) and large star-to-star variation in iron and $s$-elements (e.g. Johnson \& Pilachowski \ 2010, Marino et al.\ 2010).  Interestingly multiple populations have been detected also in the Sgr dwarf galaxy central cluster NGC 6715 (e. g. Siegel et al.\ 2007, Piotto \ 2009 and reference therein) and in Terzan 5  that is considered the surviving remnant of one of the primordial building blocks that are thought to merge and form galaxy bulges (Ferraro et al.\ 2009). 

\end{enumerate}

NGC 1851 and NGC 6656 are among the most studied clusters of the
third group and are the main subject of this paper. 
In the following we will compare the observed CMDs of these GCs and
the abundance of the critical elements Fe, O, Na, and s-elements.

\section{The color-magnitude diagram}
\begin{figure}[t!]
\resizebox{\hsize}{!}{\includegraphics[clip=true]{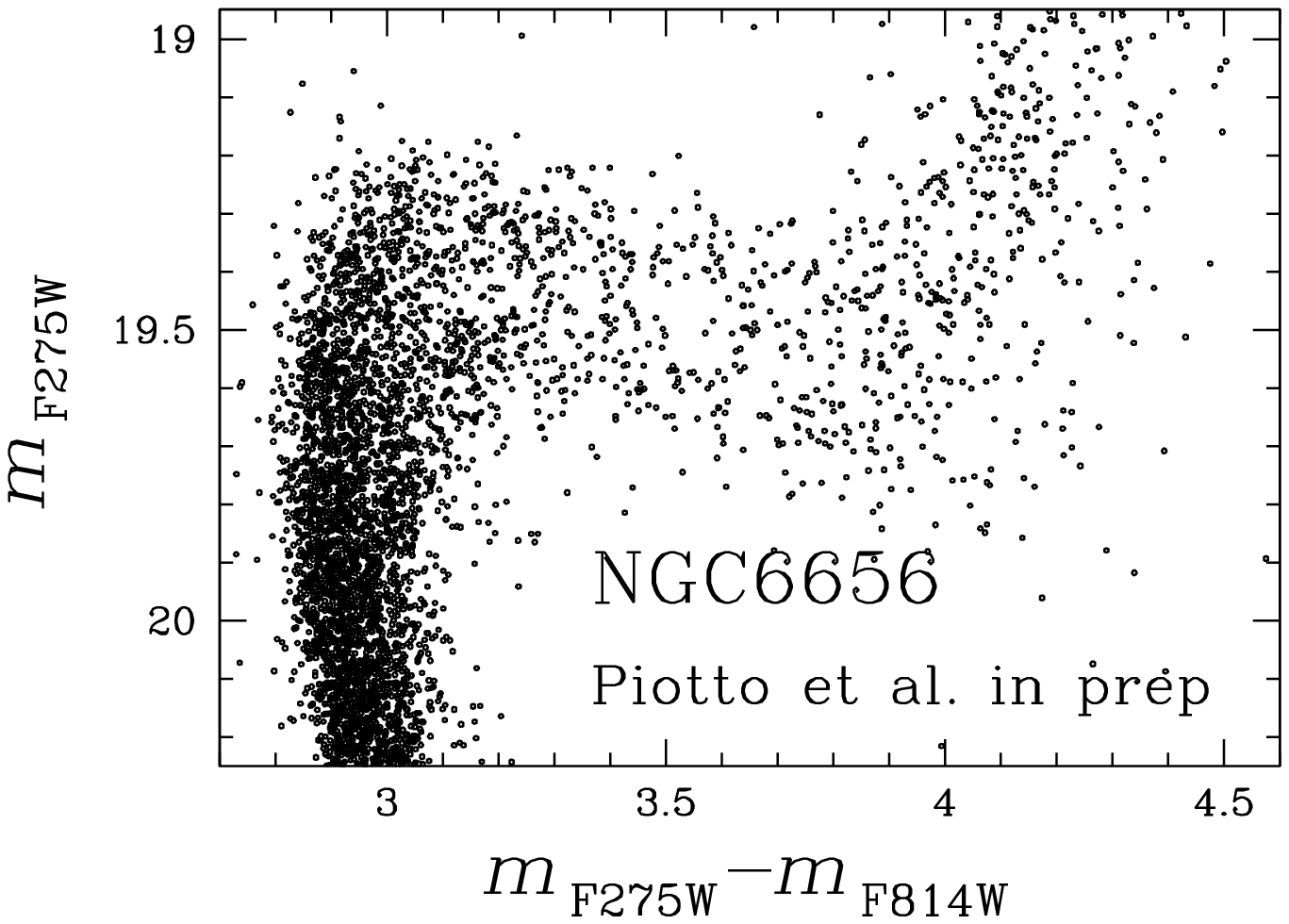}}
\resizebox{\hsize}{!}{\includegraphics[clip=true]{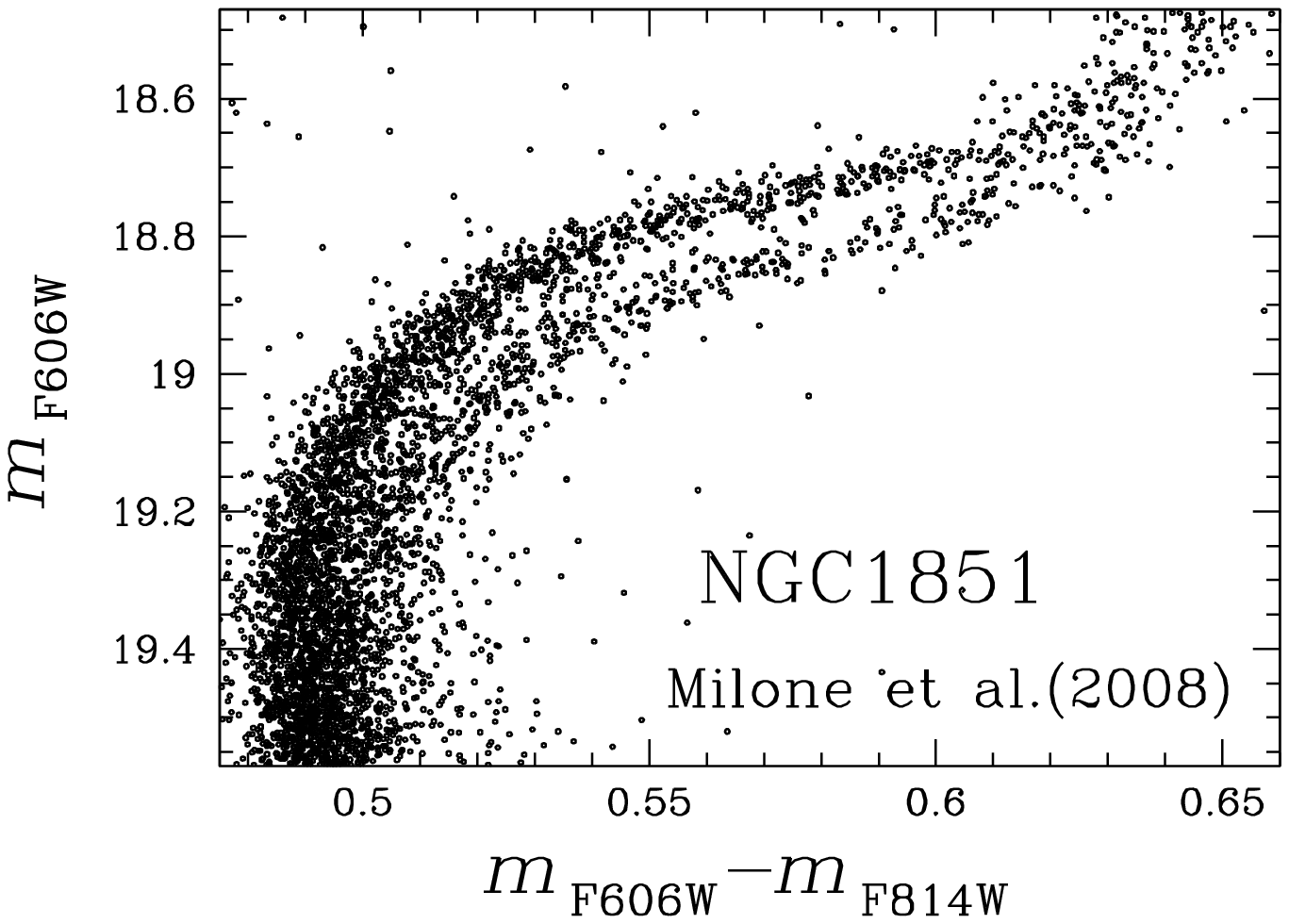}}
\caption{\footnotesize
Zoom of the CMD around the SGB for NGC 6656 ({\it upper panel}) and NGC
1851 ({\it lower panel}) from WFC3/UVIS and ACS/WFC data (GO10775, P.\ I.\ Sarajedini and GO12311, P.\ I.\ Piotto).
}
\label{CMDACS}
\end{figure}

High-accuracy photometry obtained with the WFPC2 and WFC/ACS camera on
board of the Hubble Space Telescope ({\it HST}) has revealed that the
SGB of NGC 1851 is clearly split into two branches with the bright SGB
component containing  about the 60\% of the total number of SGB
stars and the remaining 40\% of stars belonging to the faint SGB
(Milone et al.\ 2008). 
Similarly to the case of NGC 1851, after a careful correction for
differential reddening, the SGB of NGC 6656 also revealed a 
bimodality  with a fraction of bright over faint-SGB stars similar to
the one measured in NGC 1851 (Piotto \ 2009, Marino et al.\ 2009). 
A comparison of the SGBs of  NGC 1851 and NGC 6656 is given in
Fig.~\ref{CMDACS} where we show a zoom of the WFC3/UVIS CMDs around the SGB.
The spread SGB has been explained in terms of two stellar generations,
only slightly differing in age, the younger one having an increased
C+N+O abundance (Cassisi et al.\ 2008, Ventura et al.\ 2009). As an
alternative possibility, we note that observations are consistent with two stellar populations 
 with constant CNO but differing in age by $\sim$1 Gyr (Milone et al.\ 2008). 
Significant star-to-star variations in the overall CNO abundance have
been detected  among NGC 6656 RGB stars by Marino et al.\ (2010) 
and in two out four NGC 1851 giants by Yong et al. \ (2009).
The letter results is not confirmed by the recent analysis by
Villanova et al.\ (2010) who did not find significant CNO abundance variation
in 15 RGB stars of NGC 1851. 

\begin{figure}[t!]
\resizebox{\hsize}{!}{\includegraphics[clip=true]{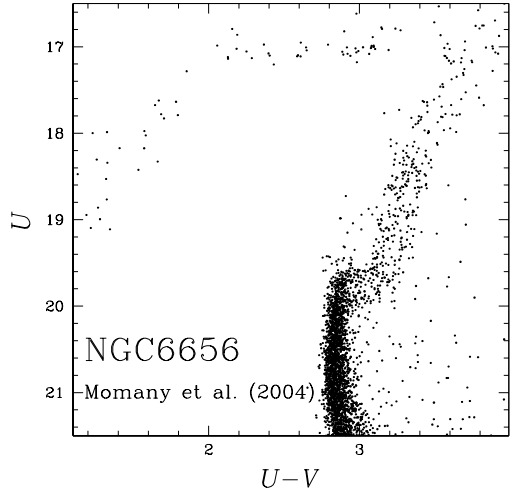}}
\caption{\footnotesize
CMD of NGC 6656 from ground-based photometry corrected for
differential reddening. 
}
\label{CMDUV}
\end{figure}

Neither the MS nor the
RGB of these clusters shows any hint of bimodality or spread in the
${\it m}_{\rm F606W}-{\it m}_{\rm F814W}$ color.
Ritcher et al.\ (1999) found that NGC 6566 exhibits a bimodal
distribution in the ${\it m}_{1}$ index among RGB stars and a similar
bimodality was detected by Calamida et al.\ (2007) for the case
of NGC 1851. A similar split among RGB stars of both clusters has been observed in the
{\it hk} index by Lee et al.\ (2009) and in the {\it U} vs. {\it (U$-$I)} and {\it U} vs. {\it (U$-$V)} CMD by Han et al.\ (2009) and Momany et al.\ (2004,
Fig.~\ref{CMDUV} ). In these cases the RGB components are clearly
associated to the two SGBs. 

\section{Chemical abundances}
\begin{figure}[t!]
\resizebox{\hsize}{!}{\includegraphics[clip=true]{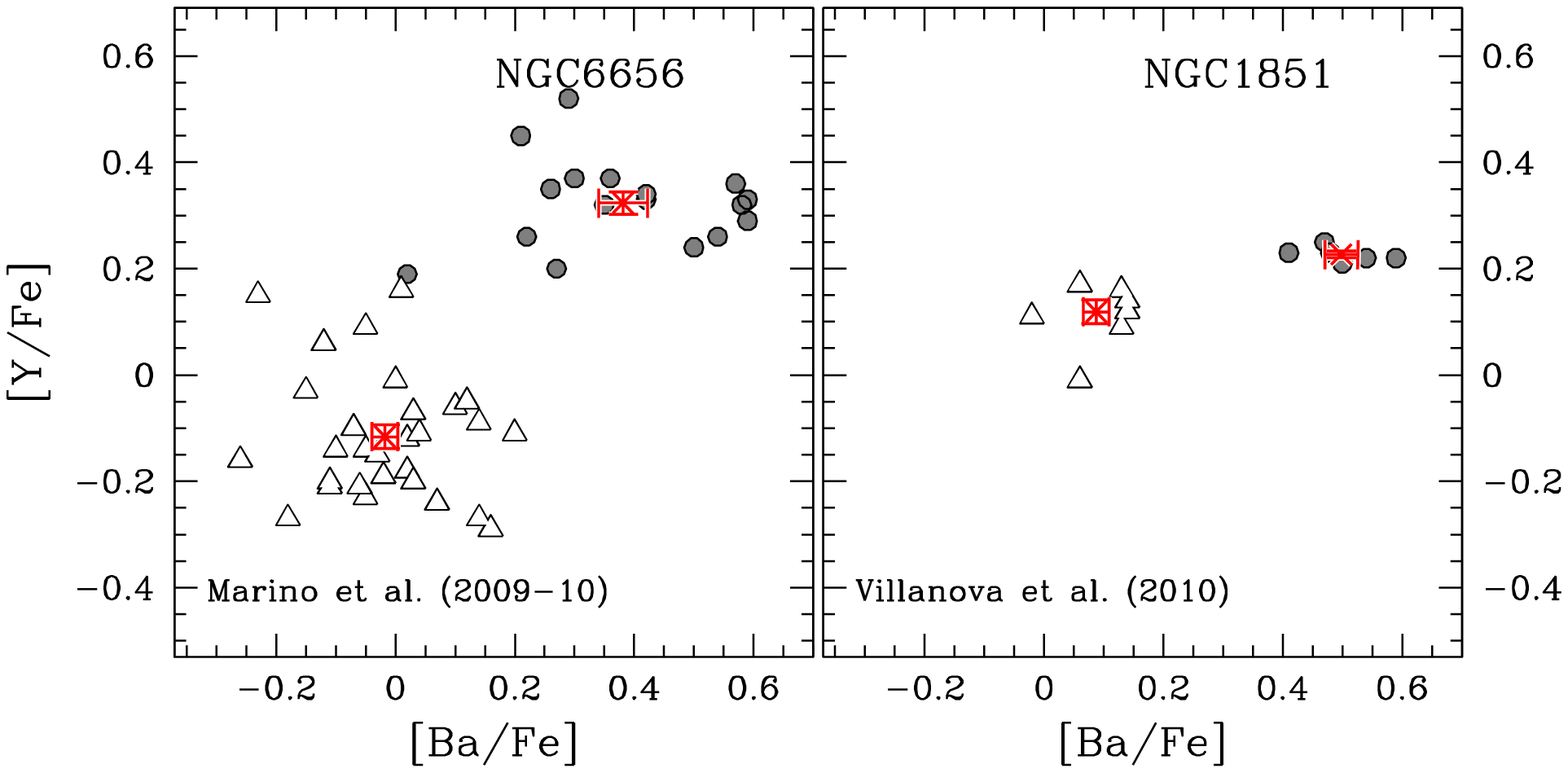}}
\resizebox{\hsize}{!}{\includegraphics[clip=true]{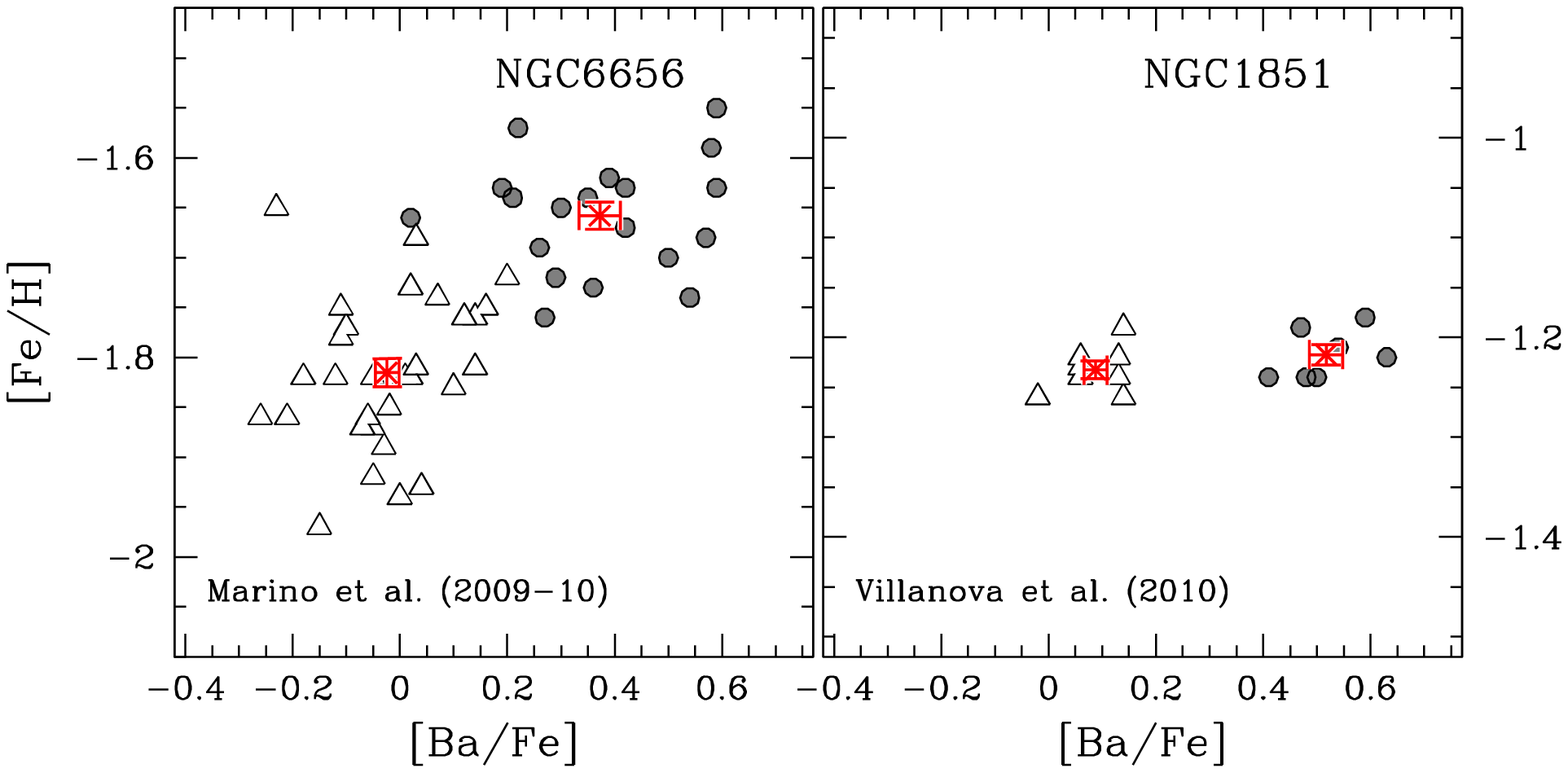}}
\caption{\footnotesize
[Y/Fe] ({\it upper panels}) and [Fe/H] ({\it lower panels}) as a
function of [Ba/Fe]. Filled circles and empty triangles represent the
$s$-rich and the $s$-poor stars. The red crosses with error bars
indicate the average abundance of stars in each group.    
}
\label{sVSs}
\end{figure}

In this section we investigate the behavior of proton-capture
elements, iron, sodium and oxygen abundances as they provide crucial
information on multiple stellar populations.
A peculiar property of the cluster pair NGC 1851-NGC 6656 is the large scatter 
 in the abundance of those $n$-capture elements that are associated to
 $s$-processes. This result comes from Marino et al.\ (2009, 2010) for
 NGC 6656 and Hesser et  al.\ (1982), Yong et al.\ (2008), and
 Villanova et al.\ (2010) for NGC 1851. As an example, Fig.~\ref{sVSs}
 shows the abundance of [Ba/Fe] as a function of [Y/Fe] for 
NGC 6656 ({\it Left Panel}) and NGC 1851 ({\it Right Panel}).  
In both clusters $n$-capture elements are clearly segregated around
two distinct values of barium and yttrium in sharp contrast from what
found in most GCs where the 
abundance of these elements does not exhibit significant star-to-star
variations. 

Since the solar-system abundances of barium and yttrium are mainly due 
to $s$-process, in the following we will call stars with high
and low [Ba/Fe] ([Y/Fe])  as `$s$-rich' and `$s$-poor'.
Following Marino et al.\ (2009), we will also use different symbols  in
Fig.~\ref{sVSs} and in the subsequent figures to separate these two groups
of stars.

When we compare the iron abundance for NGC 6656 with the $s$-elements
abundance, we find a strong correlation, with $s$-rich stars having
systematically higher [Fe/H] as shown in the lower-left panel of
Fig.~\ref{sVSs}.
The $s$-rich and $s$-poor groups have average ${\rm
  [Fe/H]_{rich}}=-1.68\pm0.02$ and ${\rm [Fe/H]_{poor}}=-1.82\pm0.02$,
respectively ($\delta {\rm [Fe/H]}=0.14\pm0.03$, Marino et
al.\ 2009). This result demonstrated that, at odds with `normal'
monometallic GCs the different stellar populations of NGC 6656 have
significant difference in their iron content. 

The presence of intrinsic spread in iron in NGC 1851 has been widely
discussed in 
literature. Yong et al.\ (2008) analyzed UVES spectra for eight RGB
stars and detected a dispersion in [Fe/H] of $\sim$0.10 dex suggesting
a possible intrinsic spread in iron. A small but detectable
metallicity spread, compatible with the presence of two groups of
stars with a metallicity difference of $\sim$0.07 dex has been claimed
also by Carretta et al.\ 2010 on the basis of their abundance analysis
of more than 120 RGB stars. These results are not confirmed 
 by Villanova et al.\ (2010) who studied high S/N MIKE/Magellan
 spectra for 15 RGB and found a smaller [Fe/H] spread compatible with
 observational errors. Villanova et al.\ (2010) detected no
 significant difference in iron between the $s$-rich and $s$-poor      
as shown in the lower-left panel of Fig.~\ref{sVSs}.

\begin{figure}[t!]
\resizebox{\hsize}{!}{\includegraphics[clip=true]{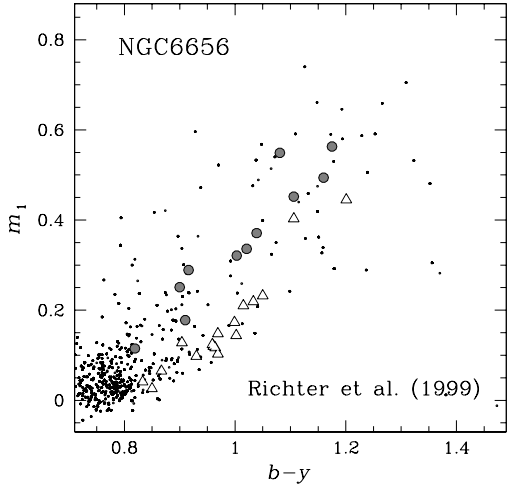}}
\caption{\footnotesize
the ${\it m}_{1}$, {\it (b-y)} diagram  for NGC 6656
corrected for differential
reddening. $s$-rich and $s$-poor stars are represented with full circles and open triangles.
}
\label{m1vscy}
\end{figure}

In the light of these results we matched spectroscopic data on NGC
6656 from Marino et al.\ (2009, 2010) with the Str\"omgren photometry by Richter et
al.\ (1999). Results are shown in Fig.~\ref{m1vscy} 
where we plotted the $m_{1}$ versus $(b-y)$ diagram for NGC 6656. 
At odds with what observed by using the {\it V} vs
{\it (V-I)} diagram, where the stars belonging to the two groups appear
to populate an unique RGB sequence, it is evident that $s$-rich and
$s$-poor stars define two distinct RGBs. 
As the $m_{1}$ index is strongly dependent by the CN bands strength,
we expect this bimodality as due to the overabundance in both C and N
measured by Marino et al.\ (2010) in $s$-rich stars. 

\subsection{Na-O anticorrelation}
\begin{figure}[t!]
\resizebox{\hsize}{!}{\includegraphics[clip=true]{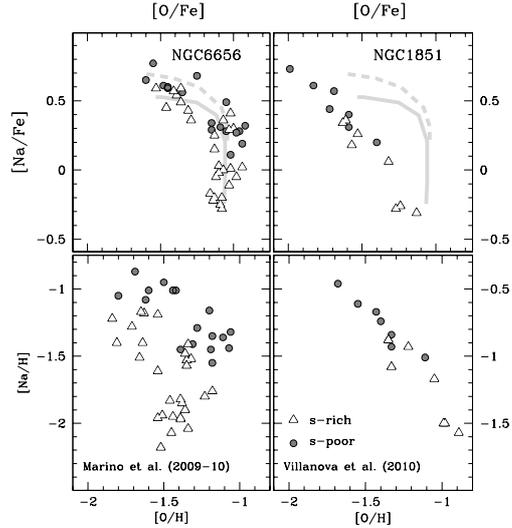}}
\caption{\footnotesize
Na-O anticorrelation for NGC 6656 ({\it upper-left panels}) and NGC 1851 ({\it
  upper-right panel}) RGB stars. $s$-rich and $s$-poor stars are
represented as full circles and open triangles.
In the upper panels the fiducial lines of the Na-O anticorrelations for the
$s$-rich and the $s$-poor stars of NGC 6656 are as continuous and
dashed gray lines respectively.  
In the lower panels we plotted [Na/H] as a function of [O/H].
}
\label{NaO}
\end{figure}

Sodium versus oxygen is plotted in Fig.~\ref{NaO} for the pair
NGC 1851-NGC 6656. 
The most important result is that in these clusters
both the $s$-rich and the $s$-poor 
group have its own Na-O anticorrelation.   
In the case of NGC 6656 also the C-N anticorrelation has been detected
in both the $s$-groups by Marino et al.\ (2010).

In both clusters  $s$-rich stars have, on average, higher sodium 
but while in the NGC 6656 both $s$-poor and $s$-rich stars span almost the same [O/Fe] range, 
in NGC 1851 $s$-rich stars are extended toward lower oxygen 
suggesting that in this case the interstellar medium could have been polluted by more massive stars.

\section{Conclusions}
We have summarized some spectroscopic and photometric findings on
multiple stellar populations in the GCs NGC 1851 and NGC 6656.

  From spectroscopic and  photometric observations, it comes out that
 the SGB and RGB split is related to the
 presence of two groups of stars with different $s$-element
 contents and a possible difference in CNO abundance. According to this
 interpretation, $s$-poor stars would constitute the first population.
 The second stellar generation should have been formed after the
 AGB winds of this first stellar generation had polluted the protocluster interstellar medium with $s$-elements. 
In the case of NGC 6656, this second generation may have formed from material that was also enriched by core-collapse supernovae ejecta, as indicated by their
 higher iron, magnesium, and silicon content, and the lack of correlation of the iron content with a pure r-process element (Eu).

This scenario is seriously challenged by the recent findings on light
elements Na, O, C, and N.
As already mentioned in Sect.~1, the Na-O and the C-N
anticorrelations can be considered the signature of multiple
star-formation episodes. 
This implies that both clusters 
are composed by two groups of stars
with distinct $s$-element content that define double SGB and RGB
observed in the CMD but each group is made by multiple
sub-populations of stars with different Na and O  abundance.
NGC 1851 and NGC 6656 do not host only two stellar
populations but have experienced a much more complex star-formation
history with similarities to the `extreme' cases of $\omega$ Centauri (Da Costa et al.\ 2009, 
Da Costa \& Marino, 2010) and the Sgr dwarf galaxy central cluster NGC 6715. 
  A detailed comparison
of the abundance patterns 
in  $\omega$ Centauri and NGC 6715 with those of NGC 1851 and 
NGC 6656 is strongly needed to resolve this problem.

\begin{acknowledgements}
We gratefully acknowledge M. Hilker, R. Kraft,  C. Sneden, and G. Wallerstein for sending us their data.
\end{acknowledgements}

\bibliographystyle{aa}

\end{document}